\documentclass[twocolumn,traditabstract]{aa}
\usepackage{txfonts}
\usepackage{color}
\usepackage{graphicx}
\usepackage{natbib}
\usepackage{multirow}
\usepackage[colorlinks=true, allcolors=blue]{hyperref}
\begin{document}

  \title{The properties of the inner disk around HL Tau: Multi-wavelength modeling of the dust emission}
   \author{Yao Liu$^{1,2,3}$
         \and
         Thomas Henning$^{1}$
         \and
         Carlos Carrasco-Gonz{\'a}lez$^{4}$
         \and
         Claire J. Chandler$^{5}$
         \and
         Hendrik Linz$^{1}$
         \and
         Til Birnstiel$^{1}$
         \and
         Roy van Boekel$^{1}$
         \and
         Laura M. P{\'e}rez$^{6}$
         \and
         Mario Flock$^{7}$
         \and
         Leonardo Testi$^{8,9,10}$
         \and 
         Luis F. Rodr{\'i}guez$^{4}$
         \and
         Roberto Galv{\'a}n-Madrid$^{4}$
         }
           \institute{Max Planck Institute for Astronomy, K\"onigstuhl 17, D-69117 Heidelberg, Germany; \\
           \email{yliu@mpia.de}
            \and
            Purple Mountain Observatory, Chinese Academy of Sciences, 2 West Beijing Road, Nanjing 210008, China
            \and
            Key Laboratory for Radio Astronomy, Chinese Academy of Sciences, 2 West Beijing Road, Nanjing 210008, China
            \and
            Instituto de Radioastronom{\'i}a y Astrof{\'i}sica UNAM, Apartado Postal 3-72 (Xangari), 58089 Morelia, Michoac{\'a}n, M{\'e}xico
            \and 
            National Radio Astronomy Observatory, P.O. Box O, 1003 Lopezville Road, Socorro, NM 87801-0387, USA
            \and
            Max Planck Institute for Radioastronomy Bonn, Auf dem H\"ugel 69, D-53121 Bonn, Germany
            \and
            Jet Propulsion Laboratory, California Institute of Technology, 4800 Oak Grove Drive, Pasadena, CA 91109, USA
            \and
            European Southern Observatory, Karl-Schwarzschild-Str. 2, D-85748 Garching bei M\"unchen, Germany
            \and
            INAF-Osservatorio Astrofisico di Arcetri, Largo E. Fermi 5, I-50125 Firenze, Italy
            \and
            Excellence Cluster ``Universe'', Boltzmann-Str. 2, D-85748 Garching bei M\"unchen, Germany}
  \authorrunning{Liu et al.}
  \titlerunning{the inner disk of HL Tau}

\abstract{We conducted a detailed radiative transfer modeling of the dust emission from the circumstellar disk around HL Tau. The goal of our study 
is to derive the surface density profile of the inner disk and its structure. In addition to the Atacama Large Millimeter/submillimeter Array images at Band 3 ($2.9\,\rm{mm}$), Band 6 ($1.3\,\rm{mm}$), 
and Band 7 ($0.87\,\rm{mm}$), the most recent Karl G. Jansky Very Large Array (VLA) observations at $7\,\rm{mm}$ were included  in the analysis. A simulated annealing algorithm 
was invoked to search for the optimum model. The radiative transfer analysis demonstrates that most radial components (i.e., $>6\,\rm{AU}$) of the disk become optically thin 
at a wavelength of 7~mm, which allows us to constrain, for the first time, the dust density distribution in the inner region of the disk. We found that a homogeneous 
grain size distribution is not sufficient to explain the observed images at different wavelengths simultaneously, while models with a shallower grain size distribution in the 
inner disk work well. We found clear evidence that larger grains are trapped in the first bright ring. Our results imply that dust evolution has already taken place in the disk at a 
relatively young (i.e., ${\sim}1\,\rm{Myr}$) age. We compared the midplane temperature distribution, optical depth, and properties of various dust rings with those reported previously. 
Using the Toomre parameter, we briefly discussed the gravitational instability as a potential mechanism for the origin of the dust clump detected 
in the first bright ring via the VLA observations.}


\maketitle

\section{Introduction}
\label{sec:intro}
HL Tau is a young (${\sim}\,1\,\rm{Myr}$) protostar located in the Taurus star formation region at a distance of ${\sim}\,140\,\rm{pc}$ \citep{rebull2004}.
A wealth of photometric measurements show strong excess emission from infrared to (sub-)millimeter wavelengths, indicating the presence of 
a substantial disk \citep{robitaille2007}. HL Tau is a Class I$-$II source based on the classification scheme of spectral 
energy distributions \citep{lada1987,white2004}. Resolved millimeter observations directly revealed a dust disk of ${\sim}150\,\rm{AU}$ in radius 
and suggested, from a comparison with the spectral energy distribution, that the millimeter grains may have settled to the midplane \citep{carrasco2009,kwon2011}. 
In addition to the disk structure, an orthogonal optical jet, a molecular bipolar outflow \citep{mundt1990}, and an 
envelope \citep{dalessio1997, menshchikov1999, robitaille2007} have been reported. Showing all ingredients of a young disk system in the 
early stage of planet formation, HL Tau is a very interesting target and has triggered extensive studies. 

Recent Atacama Large Millimeter/submillimeter Array (ALMA) observations of HL Tau revealed a pattern of bright and dark rings in the millimeter dust continuum emission 
at wavelengths of 2.9, 1.3, and $0.87\,\rm{mm}$ \citep{alma2015}. These amazing images immediately stimulated numerous 
theoretical studies. One of the most attractive scenarios is that these structures are created by embedded planets 
in the gaps, hence yielding potentially important clues regarding planet formation in such disks \citep{dong2015, dipierro2015, picogna2015}. 
Deep direct L-band imaging with the Large Binocular Telescope, however, excluded the presence of massive 
planets (${\sim}\,10\,{-}\,15\,\rm{M_{Jup}}$) in two gaps in the dust distribution at ${\sim}\,70\,\rm{AU}$ \citep{testi2015}. 
Alternative explanations for the HL Tau disk structure have also been proposed, such as the evolution of magnetized disks without planets \citep[e.g.,][]{flock2015,ruge2016},
sintering-induced dust rings \citep{okuzumi2016}, dust coagulation triggered by condensation zones of volatiles \citep{zhang2015},
and the operation of a secular gravitational instability \citep{takahashi2016}. 

\citet{pinte2016} performed detailed radiative transfer modeling of the ALMA images to account for the observed 
gaps and bright rings in the HL Tau disk. Their model shows that even at the longest ALMA wavelengths of $2.9\,\rm{mm}$, the inner disk is 
optically thick, which prevents us from correctly determining the surface density profile and the grain size distributions especially in the 
inner region of the disk where planet formation may be most efficient. Using the Karl G. Jansky Very Large Array (VLA), \citet{vla2016} 
presented the most sensitive images of HL Tau at 7 mm to date, with a spatial resolution comparable to that of the ALMA images. 
The VLA 7 mm image shows the inclined disk with a similar size and orientation as the ALMA images, and clearly identifies 
several of the features seen by \citet{alma2015}, in particular the first pair of dark (D1) and bright (B1) rings.
At this long wavelength, the dust emission from HL Tau attains a low optical depth. Assuming a temperature distribution and dust opacity,
the authors derived disk parameters such as the surface density distribution, disk mass, and individual ring masses.

Detailed radiative transfer modeling is required to constrain the temperature distribution and structural properties of 
the disk self-consistently. In this paper, we simultaneously model the ALMA and VLA images in order to reproduce the main features of the HL Tau disk
shown at different wavelengths. Our goal is to build the surface density profile, constrain the masses of individual rings, investigate 
the dust evolution, and study the gravitational stability of the disk around this fascinating object.

\section{Methodology of radiative transfer modeling}
\label{sec:modeling}

To model the dust continuum data, we use the well-tested radiative transfer code \texttt{RADMC-3D} \footnote{http://www.ita.uni-heidelberg.de/~dullemond/software/radmc-3d/.} \citep{radmc3d2012}. 
Because of the remarkable symmetry of the observed images, we assume an axisymmetric, that is, two-dimensional density structure. 
In particular, we ignore the slight offsets and change of inclinations of the rings compared to the central star \citep{alma2015}. 
Since both the ALMA and VLA images attain excellent spatial resolution and quality, our modeling is carried out directly in the image
plane. In this section, we describe the setup of the radiative transfer model and the parameters.

\subsection{Dust density distribution}
We employed a flared disk model, which has been successfully used to explain observations of similar young stars with 
circumstellar disks \citep[e.g.,][]{kenyon1987, wolfp2003, sauter2009, lium2012}. The structure of the dust density is assumed to follow a Gaussian 
vertical profile
\begin{equation}
\rho_{\rm{dust}}\propto\Sigma(R)\times\exp\left[-\frac{1}{2}\left(\frac{z}{h(R)}\right)^2\right], \\
\label{eqn:dens}
\end{equation}
where $R$ is the radial distance from the central star measured in the disk midplane, $\Sigma(R)$ is the surface density profile, and $h(R)$ is the scale 
height of the disk. The proportionality factor is determined by normalizing the mass of the entire disk. Due to the complex structures evident in the 
observations, it is not practical to directly parameterize $\Sigma(R)$ as a power law with different components. The surface density profile is built by 
adapting to the observed brightness distributions at different wavelengths (see Sect. \ref{sec:surdens}).

The disk extends from an inner radius $R_{\rm{in}}$ to an outer radius $R_{\rm{out}}$. The inner radius is set to the dust sublimation radius, that is, ${\sim}\,0.25\,\rm{AU}$. 
We fix the outer radius to 150\,AU, which is in agreement with the size of the disk estimated from resolved observations. The scale height follows a power-law profile
\begin{equation}
h(R) = H_{100}\left(\frac{R}{100\,\rm{AU}}\right)^\beta,\\
\end{equation}
with the exponent $\beta$ characterizing the degree of flaring and $H_{100}$ representing the scale height at a distance of 
$100\,\rm{AU}$ from the central star. Previous works yielded evidence of dust settling in HL Tau's disk \citep[e.g.,][]{kwon2011,pinte2016}. 
A spatial dependence of the dust properties, in particular of the grain size distribution, is required to account for all observations.
In the following, we assume that this spatial dependence is caused by dust settling in the vertical direction\footnote{Later in this paper we will also include 
the effect of radial drift of larger dust particles in the radially inward direction.} of the disk that 
is characterized by the presence of small grains in the surface layers and large grains near the midplane. We describe this dust stratification using 
a power law, with the reference scale height being a function of the grain size ($a$):
\begin{equation}
H_{100}(a) = H_{100}(a_{\rm min}) (\frac{a}{a_{\rm min}})^{-\xi}.
\label{eqn:settling}
\end{equation}
$H_{100}(a_{\rm min})$ is the reference scale height for the smallest dust grains, while the quantity $\xi$ is used to characterize the 
degree of dust settling. We fixed $\xi$ to a value of 0.1, which is typically found for other systems such as the GG Tauri circumbinary ring \citep{pinte2007} 
and IM Lupi's disk \citep{pinte2008}.
 
\begin{figure}[!t]
\includegraphics[width=0.5\textwidth]{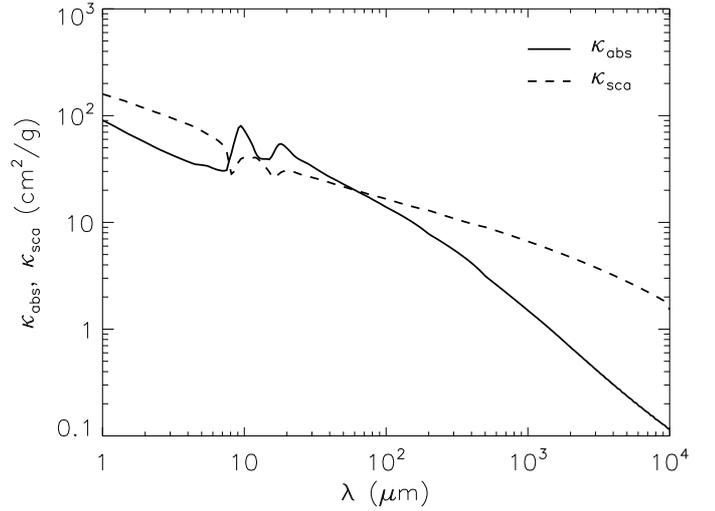}
\caption{Average mass extinction coefficients as a function of wavelength. The power-law slope $p$ of the grain size distribution is taken to be $-3.5$ for 
         this figure. The solid line stands for absorption whereas the dashed line represents scattering.}
\label{fig:kappa}
\end{figure}

\begin{figure}[!t]
\includegraphics[width=0.5\textwidth]{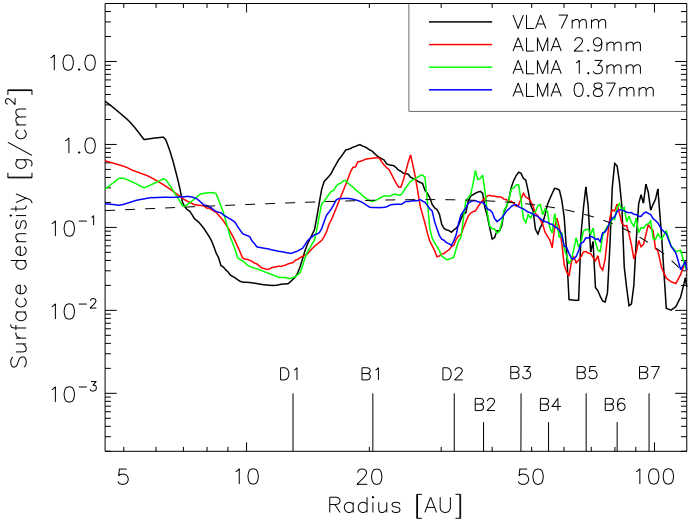}
\caption{Surface density profiles obtained from fitting the brightness distributions along the disk major axis at each wavelength (see Sect. \ref{sec:surdens}). 
The dashed line represents a tapered-edge model: $\Sigma(R) \propto (R/R_{\rm C})^{-\gamma}\, {\rm exp}[-(R/R_{\rm C})^{2-\gamma}]$ with $R_{\rm{C}}=78.9\,\rm{AU}$ 
and $\gamma=-0.22$ from \citet{kwon2011}. Selected gaps and rings identified by \citet{alma2015} are indicated. $\Sigma_{\rm VLA}$ ($7\,\rm{mm}$) is shown as black, 
$\Sigma_{\rm B3}$ ($2.9\,\rm{mm}$) as red, $\Sigma_{\rm B6}$ ($1.3\,\rm{mm}$) as green, and $\Sigma_{\rm B7}$ ($0.87\,\rm{mm}$) as blue lines.}
\label{fig:surdens}
\end{figure}

\subsection{Grain properties}
\label{sec:grainprop}
We assume that the dust grains are a mixture of 75\% amorphous silicate and 25\% carbon, with a mean density of $\rho_{\rm{grain}} = 2.5\,{\rm g\,cm^{-3}}$.
The complex refractive indices are given by \citet{dorschner1995} and \citet{jager1998}. The grain size distribution follows the standard power 
law ${\rm d}n(a)\propto{a^p} {\rm d}a$ with a minimum grain size of $a_{\rm{min}}=0.01\,\mu{\rm m}$. We fix the maximum grain size to $a_{\rm{max}}=7\,\rm{mm}$, 
which is identical to the longest wavelength of the used data sets. The observations at 7\,mm are not sensitive to larger grains. We consider 25 size bins that are logarithmically 
spaced between the minimum and maximum grain sizes. The dust extinction coefficients are computed using Mie theory. Initially, we use $p = -3.5$ (constant  
over the whole disk). However, the parameter $p$ eventually has to be modified locally due to the effects of dust evolution. Therefore, as our modeling 
progresses, we deviate from the assumption of radially constant values for $p$ over the whole disk (see Sect. \ref{sec:firstfit}). 
Figure \ref{fig:kappa} provides an illustration of the average absorption and scattering opacities over the whole grain size distribution with $p = -3.5$.

\subsection{Stellar heating}
The disk is assumed to be passively heated by stellar irradiation \citep[e.g.,][]{chiangg1997}. We take the Kurucz atmosphere models 
with ${\rm log}\,g=3.5$ as the incident stellar spectrum \citep{Kurucz1994}. The radiative transfer problem is solved self-consistently 
considering 150 wavelengths, which are logarithmically distributed in the range of [$0.1\,\mu{\rm{m}}$, $10000\,\mu{\rm{m}}$]. The effective 
temperature of the star was fixed to $4000\,\rm{K}$. For the stellar luminosity, we first adopted a value of $L \sim 6\,L_{\odot}$ as derived 
by \citet{white2004} from an analysis of HL Tau's optical spectrum. However, we found that models with this luminosity significantly underestimate 
the fluxes at all ALMA and VLA bands, even with an unreasonably large disk mass. Therefore, we increased the stellar luminosity 
to $11\,L_{\odot}$, which is identical to the value used in \citet{menshchikov1999} and \citet{pinte2016}. In \citet{white2004}, the effective 
temperature of the star was taken to be 4395\,K, whereas we adopted a value of 4000\,K. The difference in this quantity can result in different luminosities.

The disk can be actively heated by the loss of gravitational energy released from material as it accretes through the disk onto the central star
\citep[e.g.,][]{dalessio1998}, which is known as viscous heating. This heating mechanism might be important in early evolutionary 
stages (e.g., Class 0/I object), since the accretion rate is much higher than for more evolved sources. We estimated the accretion luminosity 
using the approximate formula:
\begin{equation}
L_{\rm{acc}}=\frac{G\dot{M}_{{\rm acc}}M_{\star}\left(1-\frac{R_{\star}}{R_{{\rm in}}}\right)}{R_{\star}}
\end{equation}
where $R_{\rm{in}}$ denotes the truncation radius of the inner disk and is set to be $5\,R_{\star}$ \citep{gullbring1998}.
The gravitational constant is $G$. The stellar radius ($R_{\star}$) is taken to be $7\,R_{\odot}$ and mass ($M_{\star}$) 
is $1.7\,M_{\odot}$ \citep{pinte2016}. With the accretion rate $\dot{M}_{\rm{acc}}=1.6\times10^{-7}\,M_{\odot}/yr$ measured by \citet{white2004}, 
we derived an accretion luminosity of $0.98\,L_{\odot}$ that is about $10\%$ of the assumed stellar luminosity. Furthermore, accretion 
disk theory shows that only about ${<}\,50\%$ of this accretion energy contributes to the heating of circumstellar disks \citep[e.g.,][]{calvet1998}. 
Therefore, we neglected viscous heating in our simulation.

\begin{table}[!t]
\caption{Model parameters.}
\centering
\linespread{1.5}\selectfont
\begin{tabular}{lccc}
\hline 
Parameter &   Model A    &   Model B    &  Model C \\
\hline
 \multicolumn{4}{c}{Stellar parameters} \\
\hline
$M_{\star}\, [M_{\odot}]$  & \multicolumn{3}{c}{1.7} \\
$T_{\star}\, [\rm K]$ & \multicolumn{3}{c}{4000} \\
$L_{\star}\, [L_{\odot}]$  & \multicolumn{3}{c}{11} \\
\hline
 \multicolumn{4}{c}{Disk parameters} \\
\hline
$\beta$                              &  1.15                &    $1.14^{+0.018}_{-0.007}$      &     $1.08^{+0.03}_{-0.005}$      \\
$H_{100}(a_{\rm min})\,{\rm [AU]}$   &  15                  &    $13.7^{+1.6}_{-1.0}$          &     $10.6^{+4.3}_{-1.1}$         \\ 
$M_{\rm dust}\, [10^{-3}\,M_{\odot}]$&  1.0                 &    $1.0^{+0.12}_{-0.1}$          &     $1.0^{+0.07}_{-0.16}$       \\
$p$                                  &  $-3.5$              &    --                            &     --        \\
$p_{\rm inner}$                      &  --                  &    $-3.45^{+0.08}_{-0.17}$       &     --        \\
$p_{\rm outer}$                      &  --                  &    $-3.92^{+0.18}_{-0.08}$       &     --        \\
$f_{\rm inner}$                      &  --                  &    $0.5^{+0.03}_{-0.11}$         &     --        \\
$f_{2.82}$                           &  --                  &    --                            &     $0.27^{+0.05}_{-0.07}$       \\
$f_{0.95}$                           &  --                  &    --                            &     $0.15^{+0.07}_{-0.03}$       \\
$f_{0.45}$                           &  --                  &    --                            &     $0.20^{+0.03}_{-0.06}$       \\
$f_{0.29}$                           &  --                  &    --                            &     $0.15^{+0.05}_{-0.05}$       \\
\hline 
 \multicolumn{4}{c}{Observational parameters} \\
\hline
$i\,[^{\circ}]$       &  \multicolumn{3}{c}{46.7}    \\
Position angle\,[$^{\circ}$]   & \multicolumn{3}{c}{138}   \\
$D\,{\rm [pc]}$       &  \multicolumn{3}{c}{140}  \\
\hline
\end{tabular}
\linespread{1.0}\selectfont
\tablefoot{(1) Model A is used in the derivation of the surface density profile (see Sect. \ref{sec:surdens}). 
(2) Model B and C correspond to the best-fit models under an assumption of two power-law grain size distributions (see Sect. \ref{sec:firstfit}) 
and a complex grain size distribution (see Sect. \ref{sec:secondfit}), respectively.}
\label{tab:parameter}
\end{table}

\section{Building the surface density profile}
\label{sec:surdens}
The surface density $\Sigma(R)$ is needed to complete the setup of a disk model (see Eq. \ref{eqn:dens}). It is commonly approximated by a 
simple analytic expression, in most previous studies by a power-law profile. However, ALMA observations at all three bands show a set of bright and 
dark rings in the HL Tau disk, while the VLA image unveils the structure of the first ring at a comparable spatial resolution. It is also 
obvious from these images that the gap depths (i.e., depletion factor of dust in the dark rings) are different between rings. Consequently, 
it is not practical to describe the surface density with a parameterized formula. Instead, we build the surface density of the model by 
reproducing the brightness distribution at each wavelength along the disk major axis.

We used the method introduced by \citet{pinte2016} to construct the surface density profile with an iterative procedure. 
We extracted the observed brightness profiles along the major axis of the disk in the CLEANed maps and took an average of both sides. 
We started with a simple power-law surface density with an exponent of $-0.5$. The choice of this starting profile does not really matter, 
since it is only used for the first iteration. Using the radiative transfer code \texttt{RADMC-3D}, we produced synthetic images at each wavelength 
and convolved them with appropriate Gaussian beams. When doing this, we actually simulated two images with different field of views: $30\,\rm{AU}$ and $300\,\rm{AU}$. 
Both images feature an identical number of pixels. Therefore, the image with the smaller field of view has a ten times better resolution than 
that of the other one. We combined both images into a new one after rebinning them to the pixel scale of the image with the smaller field of view. 
The extraction of model brightness distributions is performed on the new image. This procedure is employed to ensure that the radiative transfer 
models sufficiently resolve the inner disk, which is required to reproduce the central emission peak seen in both the ALMA and VLA images. 
In subsequent iterations, the surface density at each radius ($R$) is corrected by the ratio of the observed and the modeled brightness profile. 
The iteration procedure continues until the change in the model brightness profile is smaller than 2\% at any radius. It took about 35 
iterations to reach convergence and was individually performed for each of the maps.

In order to keep the procedure as simple as possible, we initially fixed $H_{100}(a_{\rm min})\,{=}\,15\,\rm{AU}$ and $\beta\,{=}\,1.15$, which are 
typically found for other disks \citep[e.g.,][]{andrews2009, madlener2012, lium2012, Kirchschlager2016}. The disk inclination ($i\,{=}\,46.7^{\circ}$) 
and position angle (${\rm PA}\,{=}\,138^{\circ}$) are adopted from the \citet{alma2015}. The slope of the grain size distribution is constant 
throughout the disk with $p\,{=}\,-3.5$. Table \ref{tab:parameter} gives an overview of the model parameters and model A corresponds to the setup 
assumed here. Figure \ref{fig:surdens} shows the result. The four surface density profiles are similar in general and we can broadly 
identify similar gap locations from them. For convenience of the subsequent parameter study, we use $\Sigma_{\rm{VLA}}$, $\Sigma_{\rm{B3}}$, 
$\Sigma_{\rm{B6}}$, and $\Sigma_{\rm{B7}}$ to label the surface density profiles that individually match the observed brightness 
distributions along the major axis of the disk at the wavelengths of 7~mm, 2.9~mm, 1.3~mm, and 0.87~mm, respectively.

\section{Modeling steps and fitting approach}

We have obtained four individual surface density profiles so far from reproducing the VLA and three ALMA images separately. The next step is to use 
one (or a combination) of them to search for a coherent model that can explain the four images simultaneously. In this section, we introduce 
the steps towards this goal and describe the approach to conduct the fitting task and the error estimate for the best-fit model. 
  
\subsection{Modeling steps}
In any modeling approach, it is preferable to build a model with as few free parameters as possible. Therefore, our modeling consists of three steps, from a 
simple scheme gradually to a more complex setup as required by the data. First, we directly take each of the four surface density profiles to check whether a radially homogeneous grain size distribution 
is sufficient to explain all the data (see Sect. \ref{sec:homsize}). Secondly, we treat the inner (${<}\,50\,\rm{AU}$) and outer (${>}\,50\,\rm{AU}$) disks separately by assuming 
different grain size distributions. We conduct a large parameter study and figure out if this new setup can reproduce all the observations (see Sect. \ref{sec:firstfit}). 
Lastly, we particularly parameterize the mass fractions of large grains with certain grain sizes and present a coherent model of HL Tau (see Sect. \ref{sec:secondfit}).

\subsection{Fitting approach}
\subsubsection{Staggered Markov chains}
\label{sec:markovchain}
We run three families of models, each of which has different degrees of freedom. The fitting task was performed with a simulated annealing (SA) 
algorithm \citep{kirkpatrick1983}, which has already been demonstrated to be successful for disk modeling \citep{madlener2012, lium2012}. Based on the 
Metropolis-Hastings algorithm, SA creates a Markov Chain Monte Carlo (MCMC), random walk through the parameter space, thereby gradually minimizing the discrepancy 
between observation and prediction by following the local topology of the merit function, for example the $\chi^2$-distribution in our case. This approach has 
specific advantages for high-dimensionality optimization because no gradients need to be calculated and entrapment in local minima can be avoided regardless 
of the dimensionality. 

Our aim is to fit all the brightness distributions in the three ALMA bands and at the VLA $7\,\rm{mm}$ band simultaneously. The traditional method used 
in multi-observable optimization is to minimize the combination of individual $\chi_{k}^2$ values from fitting each image 
\begin{equation}
\chi^2_{\rm total}=\sum_{k=1}^{N} g_{k} \, \chi^2_{k}, 
\label{eqn:chi2}
\end{equation}
which introduces arbitrary weightings ($g_{k}$) for each dataset. However, it is difficult to choose the coefficients $g_{k}$, 
because the signal-to-noise ratio (S/N) can differ between observables. In order to circumvent this problem, we devised a staggered
Markov chain \citep[for details see][]{madlener2012}. The general idea is to keep track of individual annealing temperatures of the Markov chains for 
every dataset and perform a Metropolis-Hastings decision for accepting a proposed model or rejecting it after each individual calculation 
of $\chi^2_{k}$. The 7, 2.9, 1.3, and 0.87\,mm images are fitted in a sequential order. The synthetic image at one particular wavelength is 
produced when the proposed model is accepted in the fitting of images at all longer wavelengths. By only accepting new configurations 
after all individual $\chi^2_{k}$ steps were completed successfully, the algorithm effectively optimizes all observational data sets 
simultaneously without the need for weighting them.

\subsubsection{Criteria for chain termination}
Since it is difficult to give an upper bound for the number of steps to reach the global optimum, we need to define a reasonable criterion to 
stop an optimization run. We randomly selected the starting points, propagated the chains, and monitored the quality of the fit. We found that a 
typical number of ${\sim}1000$ steps is required to achieve a ``good'' solution regardless of the starting points. This number has also been 
suggested by previous multi-wavelength modeling of circumstellar disks \citep{madlener2012}. We refrained from using a strict $\chi^2$ criterion 
to identify ``good'' solutions since we are interested here in capturing especially the qualitatively fine structure of the radial profiles 
(gaps and rings), and a model with a low $\chi^2$ does not always reproduce these features correctly.  Therefore, for each family 
of models, we calculated 15 different MCMCs starting from randomly chosen models and we let them run for 1000 steps to search for the best solution.  

\subsubsection{Error estimation}
\label{sec:error}
Estimation of the uncertainties for the best-fit model using SA is not as straightforward as in other methods like Bayesian analysis. 
First, a confidence interval $\chi^2_{\rm conf}$ has to be defined from already calculated models. Secondly, the vicinity of the best-fit 
in parameter space is probed by starting one Markov chain from this location. After collecting sufficient parameter sets below the confidence 
threshold, this normally asymmetric domain can be characterized by taking the minimum and maximum values of each degree of freedom.

We calculated a new Markov chain from the best-fit model to sample the $\chi^2$-distribution. A low chain temperature and small step widths 
for each parameter are adopted to restrain the random walk to the vicinity for sufficient steps. The total acceptance ratio of the models 
in this chain is ${\sim}\,0.2$. We gathered ∼100 accepted samples below the confidence threshold $\chi^2_{\rm conf}$ defined by 
\begin{equation}
\chi^2_{\rm conf} < 1.5 \cdot \chi^2_{\rm min},
\end{equation}
where $\chi^2_{\rm min}$ is the new overall minimum $\chi^2$ if found during the restarted run. In order to define $\chi^2_{\rm min}$ here, we have 
to use Eq. \ref{eqn:chi2} to get a compromise between the quality of the fits to each map. The coefficients $g_{k}$ are chosen in a way that 
the contributions to the $\chi^2$ come into balance between different maps. We should stress that our criterion for defining the confidence region is 
not universal due to the non-linearity of the high-dimensional models and the difference in the $\chi^2-$distribution between optimization problems.

\begin{figure}[!ht]
\includegraphics[width=0.43\textwidth]{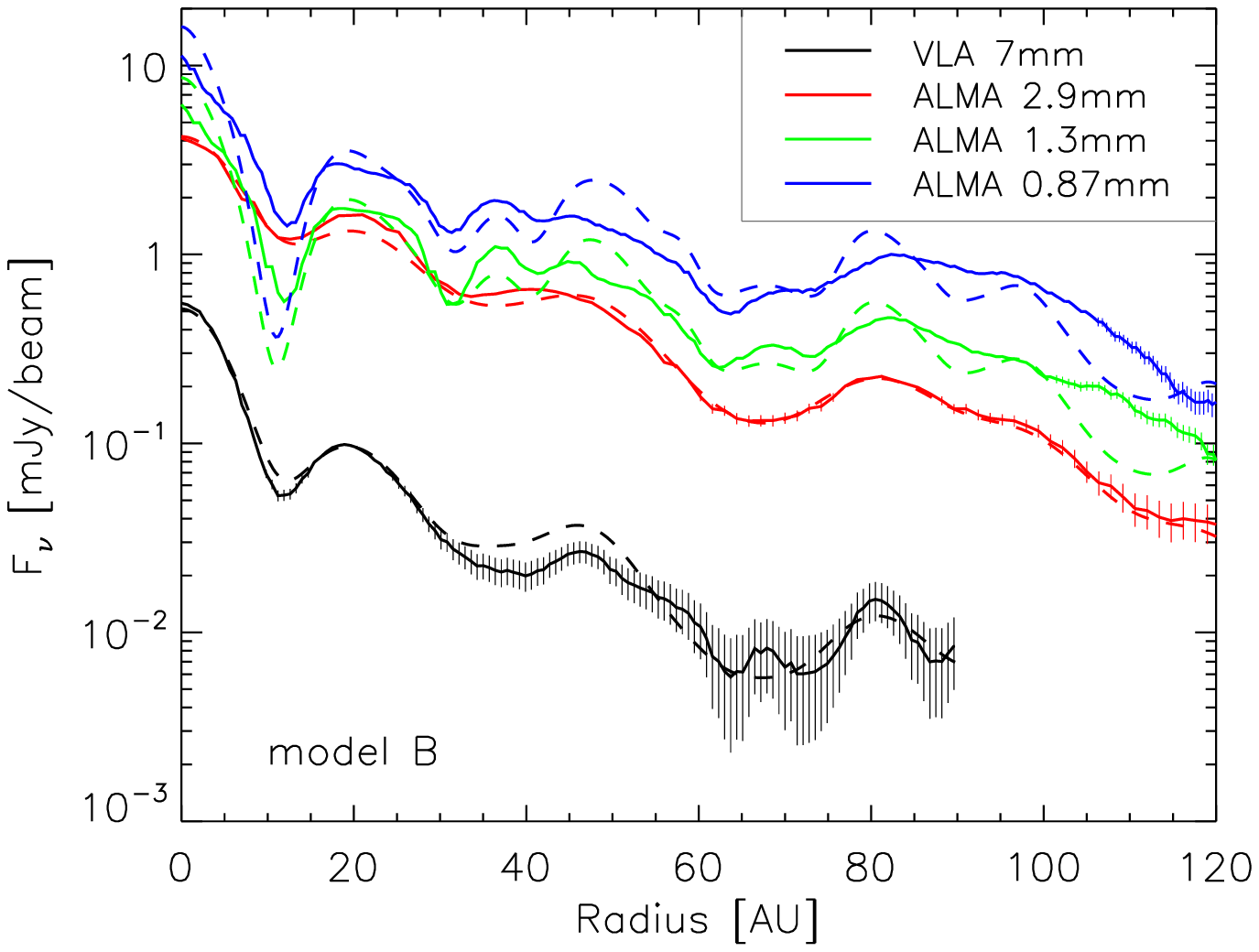}
\includegraphics[width=0.43\textwidth]{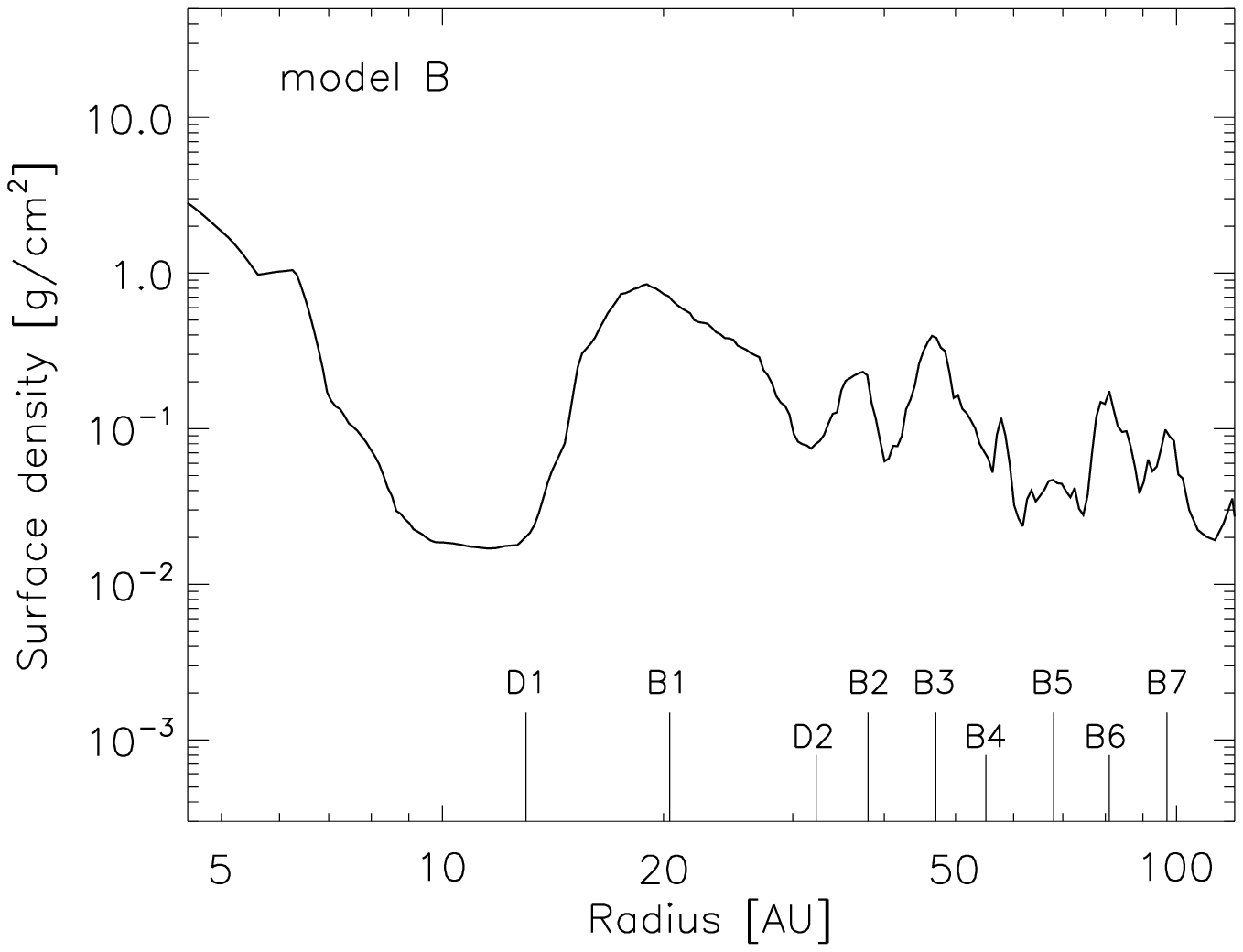}
\includegraphics[width=0.43\textwidth]{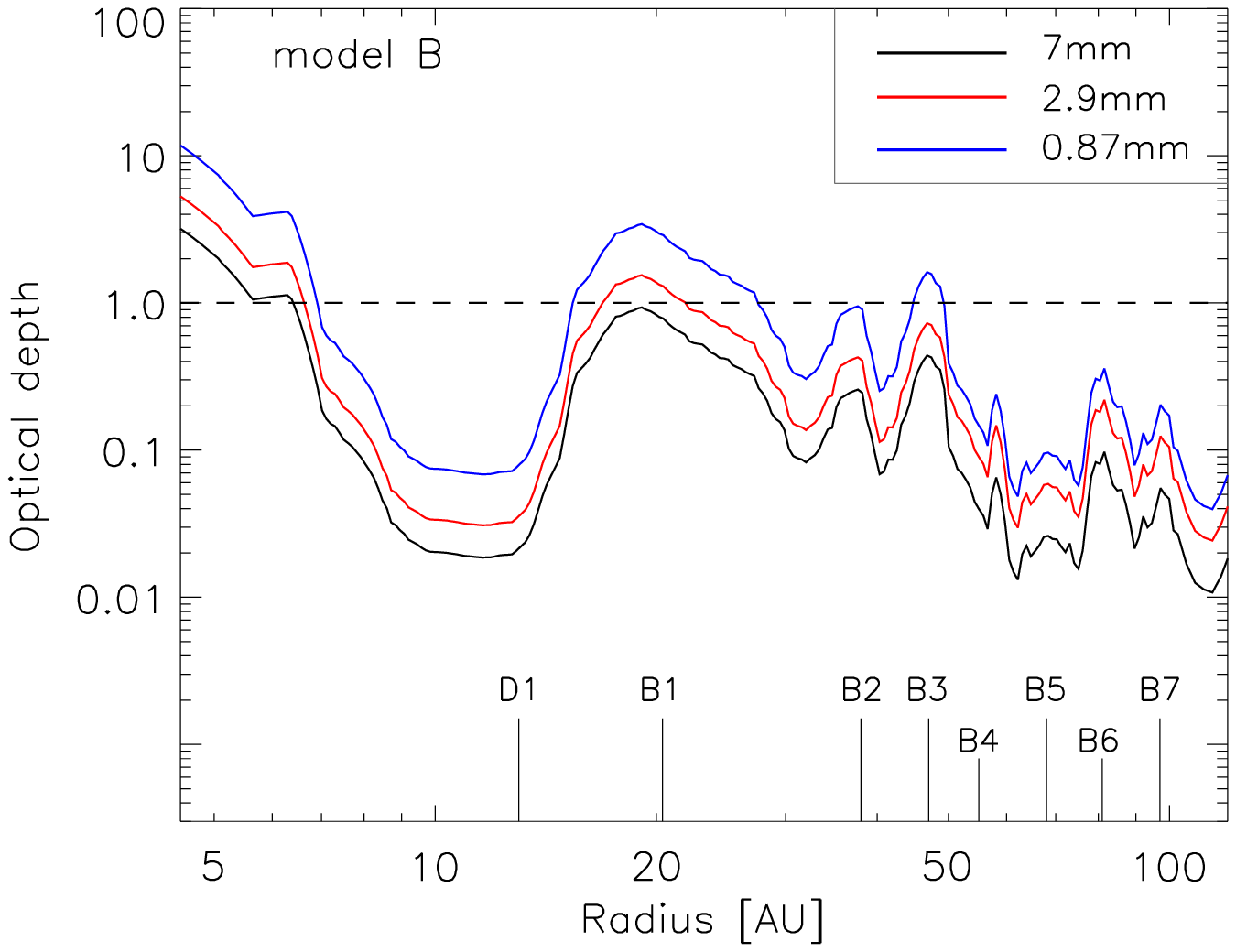}
\caption{Top panel: Observed (solid lines) and synthetic (dashed lines) flux distributions along the major axis of the disk. The results are for model B,  
in which the inner ($<50\,\rm{AU}$) and outer ($>50\,\rm{AU}$) disks have different grain size distributions (see Sect. \ref{sec:firstfit}). For better 
representation, we only show the error bars of the observations which are larger than 5\% of the fluxes. VLA $7\,\rm{mm}$ data are shown as black, 
band 3 as red, band 6 as green, and band 7 as blue lines. The VLA data at $R\,{>}\,{\sim}\,90\,\rm{AU}$ are not shown because of a low signal-to-noise ratio.
Middle panel: The combined surface density profile of model B. Bottom panel: the vertical optical depth of as a function of radius at 7, 2.9, and $0.87\,\rm{mm}$ wavelength.}
\label{fig:firstfit}
\end{figure}

\section{Results}
\subsection{Modeling with a homogeneous grain size distribution}
\label{sec:homsize}
Taking each of the four surface density profiles (i.e., $\Sigma_{\rm{VLA}}$, $\Sigma_{\rm{B3}}$, $\Sigma_{\rm{B6}}$, or $\Sigma_{\rm{B7}}$) 
as the input individually, we explored the parameter space defined as \{$\beta$, $H_{100}$, $M_{\rm dust}$, $p$\}. In other words, these parameters 
are allowed to vary freely. However, we could not find a model that reproduces all the observations simultaneously. This is an indication that 
the grain size distribution is radially changing. First hints for a radial drift of larger grains toward the inner disk have already been 
found by \citet{pinte2016} and \citet{vla2016}.

\subsection{Modeling with two power-law grain size distributions}
\label{sec:firstfit}
From a modeling analysis of HL Tau, \citet{pinte2016} found that the grain size distribution has a local slope close to $-3.5$ up to $75\,\rm{AU}$, 
while the outer disk features a steeper slope of $-4.5$, indicating that large grains in the outer disk are depleted. For optically thin 
emission in the Rayleigh-Jeans regime, the spectral index is a probe for the particle size \citep{beckwith2000}.
\citet{vla2016} calculated the spectral index $\alpha_{7-2.9\,\rm mm}$ between $7$ and $2.9\,\rm{mm}$ for the HL Tau disk. 
Their results show a clear gradient in $\alpha_{7-2.9\,\rm mm}$ between ${\sim}10$ and $50\,\rm{AU}$, with a lower value 
closer to the central star. The fact that larger grains are located at smaller radii is in agreement with general trends in theoretical 
predictions \citep{birnstiel2010,ricci2010b,ricci2010a,testi2014} and with observational constraints on several disks \citep[e.g.,][]{perez2012,perez2015}.

We therefore modify our model by introducing two new parameters, $p_{\rm inner}$ and $p_{\rm outer}$, to describe the grain size slope for the inner 
and outer disk, respectively. The two slopes are adjustable to control the contributions from large grains. Changing the maximum grain size $a_{\rm max}$ basically 
has the same effect. Therefore, we fixed $a_{\rm max}\,{=}\,7\,\rm{mm}$ in both the inner and outer disk to reduce the model degeneracy.  

Among the observations considered here, 7\,mm and 2.9\,mm are the two longest wavelengths. 
The VLA observation at 7\,mm is sensitive to the inner disk, while the structure of the outer disk is revealed by the ALMA observation. 
Therefore, for the inner 50\,AU, we take $\Sigma_{\rm{VLA}}$ as the surface density. For the outer 50\,AU, we use $\Sigma_{\rm{B3}}$. 
We chose $50\,\rm{AU}$ as the border for two reasons. First, \citet{pinte2016} show that disk regions inside this radius are (marginally) 
optically thick at the ALMA wavelengths. Therefore, the ring properties inside $50\,\rm{AU}$, derived by exclusively using the ALMA data, are 
poorly constrained. Secondly, the signal-to-noise ratio of the VLA data outside $50\,\rm{AU}$ is obviously lower compared to the inner parts.

We introduce $f_{\rm inner}$ as a control parameter on the total dust mass of the inner ($R\,{<}\,50\,\rm{AU}$) disk. The mass fraction of the outer disk 
is derived via $1-f_{\rm inner}$. The parameter space is now \{$\beta$, $H_{100}$, $M_{\rm dust}$, $p_{\rm inner}$, $p_{\rm outer}$, $f_{\rm inner}$\} (see model B in Table \ref{tab:parameter}). 
We conduct a large parameter study with the MCMC method. Figure \ref{fig:firstfit} (top panel) shows the fitting result, while the combined surface 
density profile of the best model is displayed in the middle panel. We can see that the predicted flux distributions are in good agreement 
with the observations at both 7 and 2.9\,mm. The model successfully reproduces the main features (e.g., the location and width for most dark rings) 
of the 1.3 and 0.87\,mm images as well. However, the model brightness profiles at these two wavelengths appear to have bumps and wiggles. 
In particular, the depth of the D1 ring ($R\,{\sim}\,13\,\rm{AU}$) at 1.3\,mm (band 6) and 0.87\,mm (band 7) is deeper than that of the observations. 
This is mainly due to the fact that the model does not make use of the surface density profiles $\Sigma_{\rm B6}$ and $\Sigma_{\rm B7}$, and the real 
grain size distribution in the HL Tau disk is certainly more complex than the assumed prescription. In comparison to the best-fit grain size slope 
of the inner disk $p_{\rm inner}=-3.5^{+0.1}_{-0.2}$, we observe a steeper grain size slope $p_{\rm outer}=-3.9^{+0.2}_{-0.1}$ in the outer disk, 
indicating that large grains are more concentrated in the inner region of the HL Tau disk. Figure \ref{fig:firstfit} (bottom panel) displays the 
vertical optical depth as a function of radius, from which we demonstrate that the thermal dust emission from the inner (B1 and B3) rings are 
optically thick at the ALMA wavelengths, while most parts (i.e., ${>}\,6\,\rm{AU}$) of the disk becomes optically thin at the VLA 7\,mm.

\begin{table*}[!t]
\caption{Ring properties.}
\centering
\linespread{1.5}\selectfont
\begin{tabular}{lccccc}
\hline 
Ring         &  Radius range       &   \multicolumn{3}{c}{Dust Mass  ($M_{\oplus}$)$^{2}$} \\
\cline{3-5}
             &  [AU]               & Our work   &   Carrasco-Gonzalez et al. (2016) &  Pinte et al. (2016) \\
\hline  
Peak         &  $<13$              &  $16^{+2}_{-4}$     &  10$-$50   &   $>2.3$     \\
B1           &  13$-$32            &  $43^{+2}_{-7}$     &  70$-$210  &   $>47$      \\ 
B2           &  32$-$42            &  $24^{+1}_{-4}$     &  30$-$90   &   30$-$69    \\
B3           &  42$-$50            &  $30^{+2}_{-5}$     &  20$-$80   &   14$-$37    \\
B4           &  50$-$64            &  $35^{+2}_{-6}$     &  30$-$90   &   40$-$82    \\
B5           &  64$-$74            &  $17^{+2}_{-3}$     &  10$-$50   &   5.5$-$8.7  \\
B6           &  74$-$90            &  $65^{+3}_{-9}$     &  40$-$140  &   84$-$129   \\
B7           &  90$-$150           &  $118^{+5}_{-20}$   &  --        &   99$-$157   \\
\hline
\end{tabular}
\linespread{1.0}\selectfont
\tablefoot{(1) The nomenclature used here is same as in \citet{alma2015}. (2) Uncertainties are estimated from the models with $\chi^2<1.5\cdot\chi^2_{\rm{min}}$ (see Sect. \ref{sec:error}).
(3) The ring masses are estimated based on model C (see Sect. \ref{sec:secondfit}).}
\label{tab:ringprop}
\end{table*}

\begin{figure}[!t]
\includegraphics[width=0.5\textwidth]{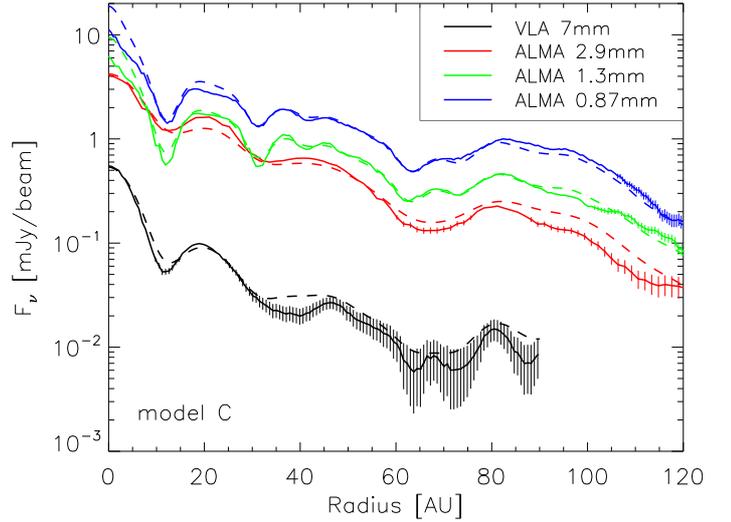}
\caption{Same as in the upper panel of Figure \ref{fig:firstfit}, but for model C (see Sect. \ref{sec:secondfit}).}
\label{fig:secondfit}
\end{figure}

\subsection{Modeling with complex grain size distributions}
\label{sec:secondfit}

By assuming different grain size slopes for the inner and outer disks, model B can explain the 7 and 2.9\,mm data well. 
However, the quality of the fit to the 1.3 and 0.87\,mm images is not very satisfying. In order to build a single model that 
reproduces the four bands simultaneously, we adopt a procedure similar to the one used in
\citet{pinte2007, pinte2016}. We utilize the four derived surface density profiles and assign each of them to a single grain size, which emits most 
at a particular wavelength. The surface densities $\Sigma_{\rm VLA}$, $\Sigma_{\rm B3}$, $\Sigma_{\rm B6}$, and 
$\Sigma_{\rm B7}$ were used to define the density distribution of the grains of size 2.82, 0.95, 0.45, and 0.29\,mm respectively. 
For all the grains with sizes in the range of [$0.01\,\mu{\rm{m}}$, 0.29\,mm], their surface densities are incorporated into $\Sigma_{\rm B7}$, independently 
of their size. Similarly, we use $\Sigma_{\rm VLA}$ to distribute all the remaining grains larger than 2.82\,mm.
Although dust of a single given grain size does not emit only at the corresponding wavelength, the contamination between bands is negligible 
because the emissivity as a function of the wavelength of a given grain size has a clear peak at these wavelengths. 

We introduce four parameters $f_{2.82}$, $f_{0.95}$, $f_{0.45}$, and $f_{0.29}$. These are mass ratios. For each of the four special grain sizes 
mentioned above, these ratio parameters denote the mass of all grains put into such a specific size  (2.82, 0.95, 0.45, and 0.29\,mm, respectively) divided 
by the entire dust mass $M_{\rm dust}$ of all grains considered (spanning a range of sizes from 0.01~$\mu$m up to 7\,mm). Therefore, the combined mass of 
grains that are either smaller than $0.29\,\rm{mm}$ or larger than $2.82\,\rm{mm}$ is thus a dependent variable through the expression 
$[(1 - (f_{2.82}+f_{0.95}+f_{0.45}+f_{0.29})] \times M_{\rm dust}$. As a result, the size distribution at any radius is no longer a simple power law. 
This new setup has a parameter space of \{$\beta$, $H_{100}$, $M_{\rm dust}$, $f_{2.82}$, $f_{0.95}$, $f_{0.45}$, $f_{0.29}$\}. 
Figure \ref{fig:secondfit} shows the best result from the MCMC optimization, whereas a summary of the model parameters (i.e., model C) can be 
found in Table \ref{tab:parameter}. The resulting model brightness distributions are now in excellent agreement with the observed ones 
at all four wavelengths. 

\begin{figure*}[!t]
\includegraphics[width=0.46\textwidth]{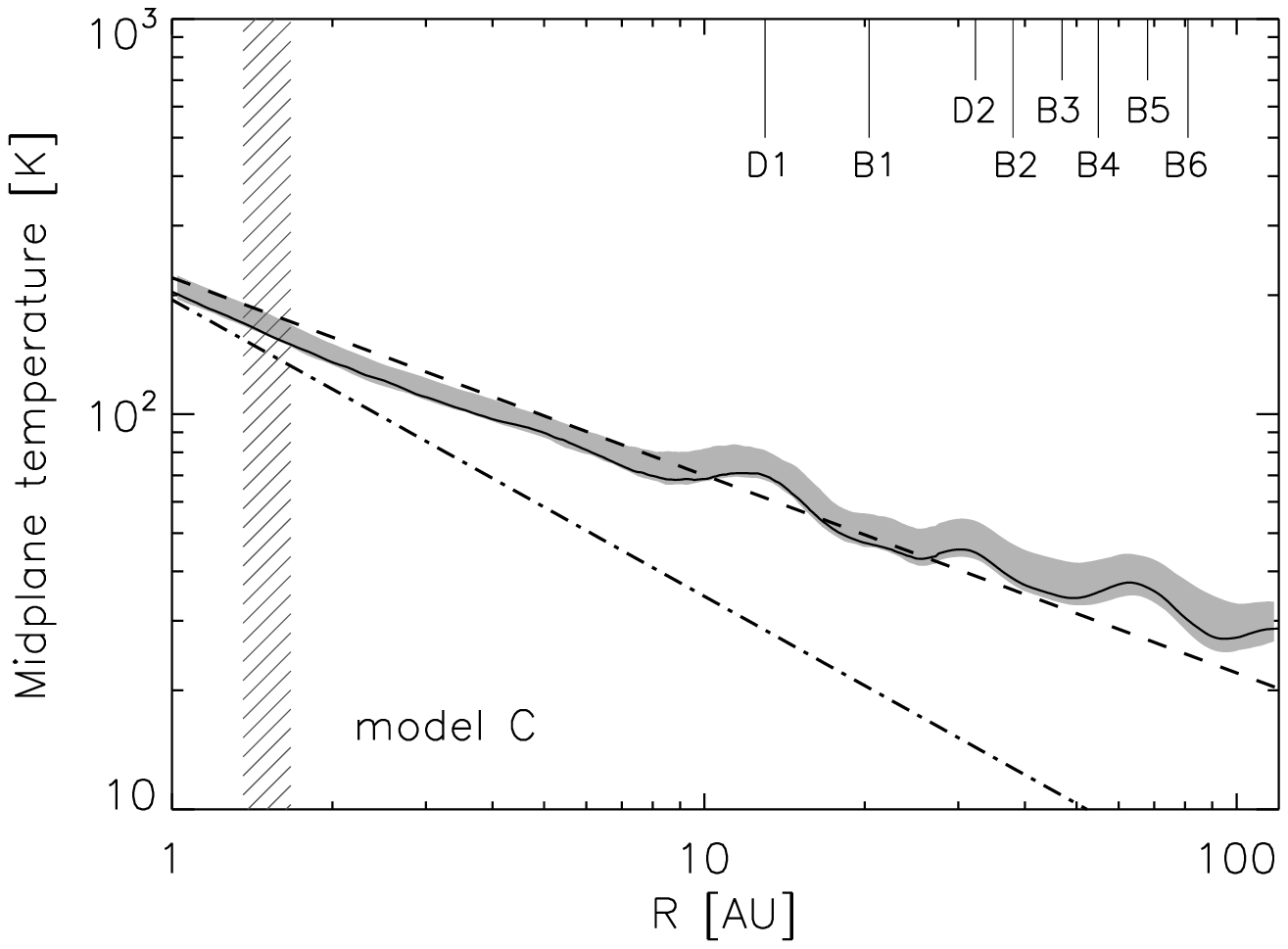}
\includegraphics[width=0.46\textwidth]{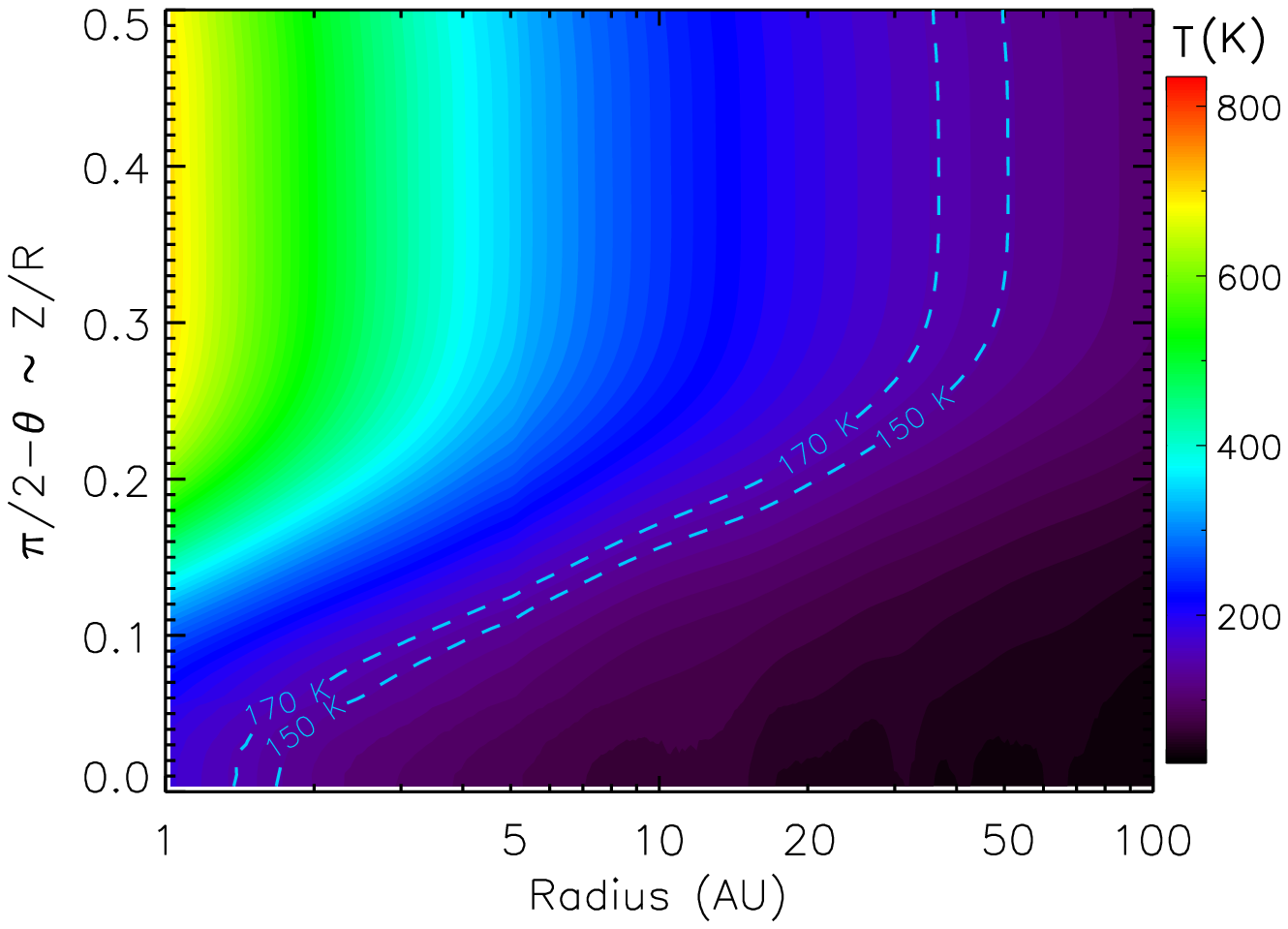}
   \caption{Left panel: Temperature distribution in the disk midplane. The black solid line refers to model C, whereas the shaded region shows the results of all 
   models within the confidence interval. The dashed line represents a power-law distribution: $T_{\rm dust}=T_{0}(R/R_{0})^{-q}$ with $q=0.5$, $T_{0}=70\,\rm{K}$ 
   and $R_{0}=10\,\rm{AU}$, which is one of the assumed temperature distributions from \citet{vla2016}. The dash-dotted line shows the effective temperature distribution of the disk 
   if disk accretion is the only heating source (see Sect. \ref{sec:innerprop}). The hashed region corresponds to the location of water snow-line ($R\,{\sim}\,1.5\,\rm{AU}$) 
   assuming an ice sublimation temperature of $150{-}170\,\rm{K}$. Right panel: The temperature structure as a function of radius $R$ and polar angle $\theta$ 
   in spherical coordinates ($\pi/2$ is the equatorial plane, i.e., the bottom of the figure) for model C. The 150\,K and 170\,K contours are indicated with dashed lines.}
\label{fig:temp}
\end{figure*}

We show the temperature distribution in the disk midplane in the left panel of Figure \ref{fig:temp}. The profile generally follows a power law. The temperature 
increases at ${\sim}\,10\,\rm{AU}$ and ${\sim}\,30\,\rm{AU}$ are presumably dominated by the indirect irradiation of the thermal emission from the B1 and B2 rings, 
while the other bump at ${\sim}\,60\,\rm{AU}$ can be explained by the density enhancement of dust grains within the B5 ring. We manually removed the B1 and B2 
rings in the simulation to test this speculation. The results show that the bumps in the midplane temperature distribution disappear when there is no bright 
irradiation source adjacent to the bump location. These fluctuations are also not caused by the noise of Monte-Carlo radiative transfer because models 
with many more photon packages show the same behavior. The right panel of Figure \ref{fig:temp} shows the two-dimensional temperature structure of the disk. 
Snow-lines of volatile species play an important role in planet formation theories \citep[e.g.,][]{pontoppidan2014}. Assuming an ice sublimation temperature 
of $150{-}170\,\rm{K}$ \citep[e.g.,][]{podolak2004}, the water snow-line is located at a radius range from ${\sim}\,1.5\,\rm{AU}$ in the midplane up 
to ${\sim}\,40\,\rm{AU}$ in the surface layers of the disk.   
 
\begin{figure}[!t]
\includegraphics[width=0.5\textwidth]{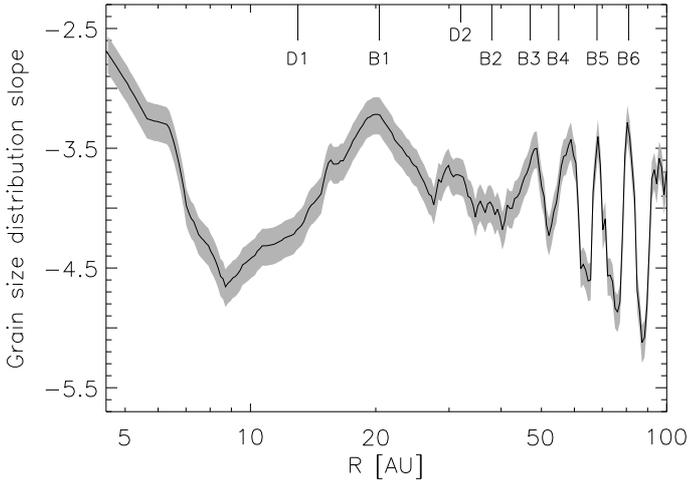}
   \caption{Slope of the local grain size distribution as a function of radius. The black solid line refers to model C, whereas the shaded region 
           shows the results of all models within the confidence interval (see Sect. \ref{sec:secondfit}).}
\label{fig:sizeslope}
\end{figure}

From the best fit, we calculate the ring properties that are summarized in Table \ref{tab:ringprop}. The nomenclature of the rings is the same as in 
\citet{alma2015}. The ring masses are calculated by integrating the surface density over the radius range between two successive gaps. In order 
to have a direct comparison with the literature results, these radius ranges are identical to those used in \citet{pinte2016} and \citet{vla2016}, which we 
list in Table \ref{tab:ringprop}. 

We convert the masses of grains with sizes of 2.82, 0.95, 0.45, and 0.29\,mm to the particle numbers $n(a)$ and vertically integrate them within 
a column ($R$). We perform a linear fit to the relation ${\rm log}\,n(a) \, vs. \, {\rm log}\,a$ to obtain the local grain size slope. 
The result is shown in Figure \ref{fig:sizeslope}. Although we see a general radial change in grain size from dominant smaller grains ($p\,{=}\,-4.5$) to 
larger grains ($p\,{=}\,-3.5$) in the inner disk, we obtain, for the first time, a spatially resolved size distribution. We clearly see that larger 
grains are accumulated in the surface density peaks, especially in the B1 ring  ($R\,{\sim}\,20\,\rm{AU}$), where the grain size distribution has 
a local slope close to $p=-3.2$ (see Figure \ref{fig:sizeslope}).


\section{Discussion}

In this section, we discuss our results concentrating on the following aspects: the properties of the inner disk, the disk 
instability, and the size distribution of dust particles.

\subsection{The properties of the inner disk}
\label{sec:innerprop}

As laid out in Sect. \ref{sec:secondfit}, we derived the masses of various rings and showed the temperature profile of the disk. According to the 
definition of the \citet{alma2015}, the gap$-$ and ring$-$like structure of the HL Tau disk is located at a radius range from ${\sim}13$ to ${\sim}90\,\rm{AU}$,
that is, from the D1 to D7 dark rings. As shown in the bottom panel of Figure \ref{fig:firstfit}, the 7\,mm dust thermal emission is optically thin 
beyond ${\sim}\,6\,\rm{AU}$, which indicates that the specific inclusion of the VLA data into the self-consistent modeling places stringent 
constraints on the properties of the disk, especially in the inner region, in contrast to the optically thick ALMA data.

The temperature profile in the disk midplane generally follows a power law, decreasing from ${\sim}\,90\,\rm{K}$ at 5\,AU to ${\sim}\,25\,\rm{K}$ at 100\,AU.
\citet{jin2016} presented a representative radiative transfer model of the HL Tau disk. In order to investigate whether the observed ring and gap features can be 
produced by planet-disk interaction, the density structure of their model is obtained from two-dimensional hydrodynamic simulations with the inclusion of three embedded 
planets and disk self-gravity. Their model features a lower midplane temperature of ${\sim}20\,\rm{K}$ at 100\,AU and a somewhat higher temperature of ${\sim}110\,\rm{K}$ at 5\,AU.
The discrepancy between our and their results is mainly due to the fact that we assumed different minimum and maximum grain sizes and dust composition. 
\citet{pinte2016} obtained a similar temperature of ${\sim}20\,\rm{K}$ at 100\,AU. \citet{vla2016} analytically estimated the total dust mass of the 
HL Tau disk by assuming a power-law temperature profile $T_{\rm dust}=T_{0}(R/R_{0})^{-q}$. In order to take the uncertainties of 
temperature into account, they explored a wide range of $q$ and $T_{0}$ in their analysis. The dashed line in Figure \ref{fig:temp} (left panel) illustrates the result 
with $q=0.5$ and $T_{0}=70\,\rm{K}$ at $R_{0}=10\,\rm{AU}$, close to the lower limit of temperatures they considered. The temperature distribution from our 
self-consistent modeling is in agreement with the temperature profile assumed by \citet{vla2016}, but in addition our results are susceptible to localized 
temperature changes in the rings. We also calculated the effective temperature of the disk when 
only viscous heating is considered as the heating process using the formula: 
\begin{equation}
T_{\rm eff, disk} = \left( \frac{3}{8\pi\sigma_{\rm{SB}}}\dot{M}_{\rm{acc}}\Omega^{2}_{K}(R) \right)^{1/4},
\end{equation} 
where $\sigma_{\rm{SB}}$ refers to the Stefan-Boltzmann constant and $\Omega_{K}(R)$ is the angular frequency \citep{knigge1999}. The result is shown as a dash-dotted
line in Figure \ref{fig:temp} (left panel). It can be seen that the effective temperatures at each radii are much lower than the midplane temperatures, which demonstrates 
again that an inclusion of viscous heating into the modeling does not significantly change the temperature structure of HL Tau's disk. 

The ring masses range from $16$ to $120\,M_{\oplus}$. Our results are generally consistent with the values given by \citet{pinte2016} and \citet{vla2016}. 
However, for individual rings there is some discrepancy probably due to the different choice of dust properties and optical depth effect affecting 
previous studies. In particular, we deduce a lower (almost two times) mass for B2 and B6 rings compared to the values presented in \citet{pinte2016}.  
In order to match the central peak emission in the ALMA images, \citet{pinte2016} fixed the surface density in the central 5\,AU to a power-law 
profile with a slope of $-0.5$ and manually added some unresolved emission at the position of the star in the simulated maps. Their results show 
that the dust emission from B2, B3, and B6 rings is marginally optically thick at 2.9\,mm, according to the criterion of $1<\tau_{2.9\,\rm{mm}}<3$.
In our approach, we produce two images with fields of view of 30 and 300\,AU at one particular wavelength. The simulated image focused on 
the central region has a better resolution and can help to get close to the correct flux prediction. By conducting a large parameter study, the 
optimization algorithm tends to put more dust near the center to reproduce the peaks, but without any further addition of emission. This is the reason 
why a clear increase appears in the surface density profile towards the central star (see the black line in Figure \ref{fig:surdens}). 
As a consequence, the outer region where these rings are located shares less of the disk's mass. From the bottom panel of Figure \ref{fig:firstfit}, 
one can see that the dust thermal emission from the B2, B3, and B6 rings is optically thin at both 2.9 and 7\,mm. Hence, the inclusion of the 
VLA observation into the modeling analysis clearly improves the constraint on the ring properties. With the dust properties described in 
Sect. \ref{sec:grainprop}, model C has an average (averaged over the whole radius range) dust mass extinction coefficient at 7\,mm of 
$\kappa_{\rm{abs}}{+}\kappa_{\rm{sca}}=1.6\,\rm{cm^2/g}$, which is clearly larger 
than the maximum value of $0.2\,\rm{cm^2/g}$ considered in \citet{vla2016}. As a result, we obtained systematically lower ring masses than their results.

\begin{figure}[!t]
\includegraphics[width=0.5\textwidth]{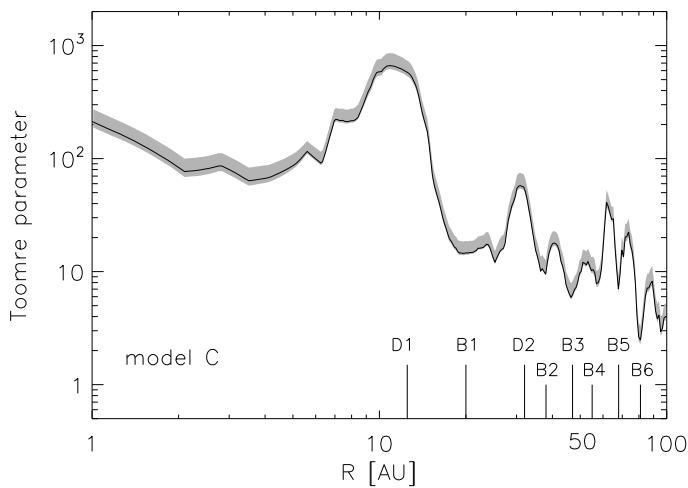}
   \caption{Toomre parameter as a function of radius. The black solid line refers to model C, whereas the shaded region shows the 
            results from all the models within the confidence interval.}
\label{fig:toomre}
\end{figure}

\subsection{Disk instability}

The highest angular resolution VLA images of the HL Tau disk reveal a compact knot structure in the first bright ring (B1) \citep{vla2016}.
This knot also coincides with a local intensity maximum in the ALMA 1.3\,mm image. One of the formation mechanisms of a clumpy structure  
is gravitational instability (GI) of massive disks.

We checked this possibility using the Toomre criterion (Toomre 1964) that describes when a disk becomes unstable through gravitational collapse. 
The Toomre $Q$ parameter is defined as
\begin{equation}
Q=\frac{c_{s}\Omega_{k}}{\pi G \Sigma},
\end{equation}
where $c_s$ is the local sound speed and $\Sigma$ stands for the surface density of the gas. In terms of $Q$, a disk is unstable to its own 
self-gravity if $Q < 1$, and stable if $Q > 1$. Figure \ref{fig:toomre} displays the Toomre parameter as a function of radius. We assumed a 
constant dust-to-gas mass ratio of 0.01 throughout the disk for simplicity. However, \citet{yen2016} analyzed the ${\rm HCO^{+}(1-0)}$ data 
of HL Tau obtained from ALMA. Their results indicate that the dust-to-gas mass ratio probably varies as a function of radius. In particular, 
they found a relatively higher fraction of gas in the first dark (D1) ring. Taking this fact into account, the prominent bump of $Q$ at 
the location of the D1 ring will be reduced (see Figure \ref{fig:toomre}).

It is clear that the $Q$ value is greater than unity for the entire disk. However, for $R\,{>}\,80\,\rm{AU}$, the disk is close to the threshold of 
being gravitationally unstable. This is consistent with the recent finding by \citet{akiyama2016} that the GI can operate most probably in the outer 
region of the HL Tau disk. \citet{pohl2015} showed that an embedded massive planet (planet-to-star mass ratio of 0.01) can trigger a gravitational 
instability in massive disks with a Toomre parameter around unity. Assuming a stellar mass of $1.7\,M_{\odot}$, a planet with a mass of 
${>}18\,M_{\rm{Jup}}$ is required. With direct mid-IR imaging, \citet{testi2015} set an upper limit of ${\sim}10{-}15\,M_{\rm{Jup}}$ 
(i.e., smaller than the required mass) for planetary objects present in the D5 and D6 dark rings. However, massive planets may exist in other locations, 
and the effect of multiple (small mass) planets on the reduction of the Toomre parameter has not yet been investigated. 

Another possibility for a driver of structure formation, however, is secular GI.
Secular GI is the gravitational collapse of dust due to gas$-$dust friction \citep[e.g.,][]{ward2000,youdin2011}. Recently, secular GI has been proposed 
as an underlying mechanism for multiple ring$-$like structures \citep{takahashi2014}. Using physical parameters derived from the ALMA observations, 
\citet{takahashi2016} performed a linear stability analysis of secular GI, and verified that it can cause the ring structures observed in HL Tau
under certain conditions. The assumed temperature distribution, surface density, and maximum grain size (as the main inputs of the secular GI analysis) 
agree well with what we obtained from modeling the VLA and ALMA observations simultaneously. Since secular GI can concentrate the dust and 
support dust growth, the observed dust clump may be a consequence of dust condensation in the first bright ring (B1) induced by secular GI.

\subsection{Grain size distribution}
The growth of dust grains from sub$-$$\mu\rm{m}$ to millimeter sizes is the first step towards the formation of planetesimals \citep[e.g.,][]{papaloizou2006,lissauer2007,natta2007}.
\citet{brauer2008} and \citet{birnstiel2010a} have modeled the evolution of dust grain populations with a sophisticated treatment of the dust growth processes and the 
dynamical mechanisms responsible for the transport of dust grains in disks. These theoretical studies predict that dust properties change as a function of the radius, 
with more large dust grains located at smaller radii \citep[see also][]{birnstiel2010}. 

The (sub-)millimeter spectral slope ($\alpha$) derived from spatially resolved observations is a common probe for grain size: low values of $\alpha$ can be interpreted in 
terms of grain growth \citep[e.g.,][]{isella2010,guilloteau2011}. Radius-dependent dust properties have been observed for a number of disks, for example
AS 209 \citep{perez2012}, CY Tau, and DoAr 25 \citep{perez2015}. Moreover, \citet{trotta2013} and \citet{tazzari2016} carried out self-consistent modeling works 
of the disk structure and the dust properties by fitting interferometric observations of a sample of targets using a radius-dependent grain size distribution.
Their results confirm the radial variation of grain growth, with the maximum grain size decreasing with radius. 

\citet{vla2016} calculated the spectral index $\alpha_{7-2.9\,\rm{mm}}$ between 7 and 2.9\,mm, which are currently the two most optically thin wavelengths of the 
observations for HL Tau. The authors witnessed a clear radial gradient in $\alpha_{7-2.9\,\rm{mm}}$ that is consistent with a change in dust properties and a 
differential grain size distribution, with more larger grains at smaller radii. This is confirmed by our modeling. Model B presented in Sect. \ref{sec:firstfit} 
has a grain size slope of $p_{\rm inner}=-3.5^{+0.1}_{-0.2}$ in the inner disk and a steeper value of $p_{\rm outer}=-3.9^{+0.2}_{-0.1}$ 
for the outer ($R\,{>}\,50\,\rm{AU}$) disk. \citet{pinte2016} also obtained a shallower slope at smaller radii from modeling the ALMA data. 
Model C shows the same general trend, with $p$ approaching values of $-4.5$ beyond 60\,AU (see Sect. \ref{sec:secondfit}).  
In addition, model C reveals that the slope for the grain size distribution has local maxima associated with many of the bright rings. For instance, the grain size distribution of the B1 ring 
($R\,{\sim}\,20\,\rm{AU}$) has a local slope close to $p=-3.2$  This can be interpreted as an indication for a concentration of larger grains in such regions. 
All this suggests that dust growth and concentration of large grains, as initial steps of planet formation, have 
already occurred in HL Tau at the relatively young age of ${\sim}1\,\rm{Myr}$.

\section{Summary}
We performed detailed radiative transfer modeling of the HL Tau disk. Both the ALMA images at 0.87, 1.3, and 2.9\,mm and the recent VLA 7\,mm map 
are taken into account. We first built the surface density profiles by inverting the observed brightness distributions along the major axis of the 
disk at each individual wavelength. Then, a large parameter study was conducted to search for a single model that can interpret all observations simultaneously. 
In order to reduce the model degeneracy as far as possible, our modeling starts from a simple assumption with a radially homogeneous grain size distribution, 
moving to a more complex scenario where the inner and outer disks feature different grain size slopes, to a prescription where the mass fractions of large dust grains 
are parameterized. We are able to place better constraints on the inner disk structure and the properties of different rings than all results 
exclusively based on ALMA observations because we added the more optically thin VLA 7~mm data to our analysis.

The temperature distribution in the disk midplane generally follows a power law, decreasing from ${\sim}\,90\,\rm{K}$ at 5\,AU to ${\sim}\,25\,\rm{K}$ at 100\,AU.
The dust masses in the rings range from $16$ to $120\,M_{\oplus}$. We found a shallower slope of grain sizes in the inner disk, which indicates that a higher 
fraction of larger grains is present closer to the star. This is consistent with theoretical predictions of circumstellar disk evolution. In particular, many 
of the bright rings are associated with local maxima of the grain size distribution slope, indicating a concentration of larger grains in those rings. The Toomre 
parameter $Q$ is greater than unity for the entire disk. However, for $R\,{>}\,80\,\rm{AU}$, the disk is close to the threshold of being gravitationally unstable. 
We also discussed the origins of the dust clump in the first bright ring revealed by the VLA and suggested that secular GI is a promising mechanism. 

\acknowledgements
We thank the anonymous referees for constructive comments that improved the manuscript. We thank Dr.~Adriana Pohl, Dr.~Yuan Wang, and Dr.~Joel Sanchez-Bermudez for valuable discussions. 
Y.L. acknowledges supports by NSFC grant 11503087 and by the Natural Science Foundation of Jiangsu Province of China (Grant No. BK20141046). Y.L. acknowledges supports 
by the German Academic Exchange Service and the China Scholarship Council. T.B. acknowledges support from the DFG through SPP 1833 "Building a Habitable Earth" (KL 1469/13-1).

\bibliographystyle{aa}
\bibliography{hltau}

\begin{thebibliography}{65}
\expandafter\ifx\csname natexlab\endcsname\relax\def\natexlab#1{#1}\fi

\bibitem[{{Akiyama} {et~al.}(2016){Akiyama}, {Hasegawa}, {Hayashi}, \&
  {Iguchi}}]{akiyama2016}
{Akiyama}, E., {Hasegawa}, Y., {Hayashi}, M., \& {Iguchi}, S. 2016, \apj, 818,
  158

\bibitem[{{ALMA Partnership} {et~al.}(2015){ALMA Partnership}, {Brogan},
  {P{\'e}rez}, {Hunter}, {Dent}, {Hales}, {Hills}, {Corder}, {Fomalont},
  {Vlahakis}, {Asaki}, {Barkats}, {Hirota}, {Hodge}, {Impellizzeri}, {Kneissl},
  {Liuzzo}, {Lucas}, {Marcelino}, {Matsushita}, {Nakanishi}, {Phillips},
  {Richards}, {Toledo}, {Aladro}, {Broguiere}, {Cortes}, {Cortes}, {Espada},
  {Galarza}, {Garcia-Appadoo}, {Guzman-Ramirez}, {Humphreys}, {Jung}, {Kameno},
  {Laing}, {Leon}, {Marconi}, {Mignano}, {Nikolic}, {Nyman}, {Radiszcz},
  {Remijan}, {Rod{\'o}n}, {Sawada}, {Takahashi}, {Tilanus}, {Vila Vilaro},
  {Watson}, {Wiklind}, {Akiyama}, {Chapillon}, {de Gregorio-Monsalvo}, {Di
  Francesco}, {Gueth}, {Kawamura}, {Lee}, {Nguyen Luong}, {Mangum}, {Pietu},
  {Sanhueza}, {Saigo}, {Takakuwa}, {Ubach}, {van Kempen}, {Wootten},
  {Castro-Carrizo}, {Francke}, {Gallardo}, {Garcia}, {Gonzalez}, {Hill},
  {Kaminski}, {Kurono}, {Liu}, {Lopez}, {Morales}, {Plarre}, {Schieven},
  {Testi}, {Videla}, {Villard}, {Andreani}, {Hibbard}, \&
  {Tatematsu}}]{alma2015}
{ALMA Partnership}, {Brogan}, C.~L., {P{\'e}rez}, L.~M., {et~al.} 2015, \apjl,
  808, L3

\bibitem[{{Andrews} {et~al.}(2009){Andrews}, {Wilner}, {Hughes}, {Qi}, \&
  {Dullemond}}]{andrews2009}
{Andrews}, S.~M., {Wilner}, D.~J., {Hughes}, A.~M., {Qi}, C., \& {Dullemond},
  C.~P. 2009, \apj, 700, 1502

\bibitem[{{Beckwith} {et~al.}(2000){Beckwith}, {Henning}, \&
  {Nakagawa}}]{beckwith2000}
{Beckwith}, S.~V.~W., {Henning}, T., \& {Nakagawa}, Y. 2000, Protostars and
  Planets IV, 533

\bibitem[{{Birnstiel} {et~al.}(2010{\natexlab{a}}){Birnstiel}, {Dullemond}, \&
  {Brauer}}]{birnstiel2010a}
{Birnstiel}, T., {Dullemond}, C.~P., \& {Brauer}, F. 2010{\natexlab{a}}, \aap,
  513, A79

\bibitem[{{Birnstiel} {et~al.}(2010{\natexlab{b}}){Birnstiel}, {Ricci},
  {Trotta}, {Dullemond}, {Natta}, {Testi}, {Dominik}, {Henning}, {Ormel}, \&
  {Zsom}}]{birnstiel2010}
{Birnstiel}, T., {Ricci}, L., {Trotta}, F., {et~al.} 2010{\natexlab{b}}, \aap,
  516, L14

\bibitem[{{Brauer} {et~al.}(2008){Brauer}, {Dullemond}, \&
  {Henning}}]{brauer2008}
{Brauer}, F., {Dullemond}, C.~P., \& {Henning}, T. 2008, \aap, 480, 859

\bibitem[{{Calvet} \& {Gullbring}(1998)}]{calvet1998}
{Calvet}, N. \& {Gullbring}, E. 1998, \apj, 509, 802

\bibitem[{{Carrasco-Gonz{\'a}lez} {et~al.}(2016){Carrasco-Gonz{\'a}lez},
  {Henning}, {Chandler}, {Linz}, {P{\'e}rez}, {Rodr{\'{\i}}guez},
  {Galv{\'a}n-Madrid}, {Anglada}, {Birnstiel}, {van Boekel}, {Flock}, {Klahr},
  {Macias}, {Menten}, {Osorio}, {Testi}, {Torrelles}, \& {Zhu}}]{vla2016}
{Carrasco-Gonz{\'a}lez}, C., {Henning}, T., {Chandler}, C.~J., {et~al.} 2016,
  \apjl, 821, L16

\bibitem[{{Carrasco-Gonz{\'a}lez} {et~al.}(2009){Carrasco-Gonz{\'a}lez},
  {Rodr{\'{\i}}guez}, {Anglada}, \& {Curiel}}]{carrasco2009}
{Carrasco-Gonz{\'a}lez}, C., {Rodr{\'{\i}}guez}, L.~F., {Anglada}, G., \&
  {Curiel}, S. 2009, \apjl, 693, L86

\bibitem[{{Chiang} \& {Goldreich}(1997)}]{chiangg1997}
{Chiang}, E.~I. \& {Goldreich}, P. 1997, \apj, 490, 368

\bibitem[{{D'Alessio} {et~al.}(1997){D'Alessio}, {Calvet}, \&
  {Hartmann}}]{dalessio1997}
{D'Alessio}, P., {Calvet}, N., \& {Hartmann}, L. 1997, \apj, 474, 397

\bibitem[{{D'Alessio} {et~al.}(1998){D'Alessio}, {Cant{\"o}}, {Calvet}, \&
  {Lizano}}]{dalessio1998}
{D'Alessio}, P., {Cant{\"o}}, J., {Calvet}, N., \& {Lizano}, S. 1998, \apj,
  500, 411

\bibitem[{{Dipierro} {et~al.}(2015){Dipierro}, {Price}, {Laibe}, {Hirsh},
  {Cerioli}, \& {Lodato}}]{dipierro2015}
{Dipierro}, G., {Price}, D., {Laibe}, G., {et~al.} 2015, \mnras, 453, L73

\bibitem[{{Dong} {et~al.}(2015){Dong}, {Zhu}, \& {Whitney}}]{dong2015}
{Dong}, R., {Zhu}, Z., \& {Whitney}, B. 2015, \apj, 809, 93

\bibitem[{{Dorschner} {et~al.}(1995){Dorschner}, {Begemann}, {Henning},
  {Jaeger}, \& {Mutschke}}]{dorschner1995}
{Dorschner}, J., {Begemann}, B., {Henning}, T., {Jaeger}, C., \& {Mutschke}, H.
  1995, \aap, 300, 503

\bibitem[{{Dullemond} {et~al.}(2012){Dullemond}, {Juhasz}, {Pohl}, {Sereshti},
  {Shetty}, {Peters}, {Commercon}, \& {Flock}}]{radmc3d2012}
{Dullemond}, C.~P., {Juhasz}, A., {Pohl}, A., {et~al.} 2012, {RADMC-3D: A
  multi-purpose radiative transfer tool}, Astrophysics Source Code Library

\bibitem[{{Flock} {et~al.}(2015){Flock}, {Ruge}, {Dzyurkevich}, {Henning},
  {Klahr}, \& {Wolf}}]{flock2015}
{Flock}, M., {Ruge}, J.~P., {Dzyurkevich}, N., {et~al.} 2015, \aap, 574, A68

\bibitem[{{Guilloteau} {et~al.}(2011){Guilloteau}, {Dutrey}, {Pi{\'e}tu}, \&
  {Boehler}}]{guilloteau2011}
{Guilloteau}, S., {Dutrey}, A., {Pi{\'e}tu}, V., \& {Boehler}, Y. 2011, \aap,
  529, A105

\bibitem[{{Gullbring} {et~al.}(1998){Gullbring}, {Hartmann}, {Briceno}, \&
  {Calvet}}]{gullbring1998}
{Gullbring}, E., {Hartmann}, L., {Briceno}, C., \& {Calvet}, N. 1998, \apj,
  492, 323

\bibitem[{{Isella} {et~al.}(2010){Isella}, {Carpenter}, \&
  {Sargent}}]{isella2010}
{Isella}, A., {Carpenter}, J.~M., \& {Sargent}, A.~I. 2010, \apj, 714, 1746

\bibitem[{{J\"ager} {et~al.}(1998){J\"ager}, {Mutschke}, \&
  {Henning}}]{jager1998}
{J\"ager}, C., {Mutschke}, H., \& {Henning}, T. 1998, \aap, 332, 291

\bibitem[{{Jin} {et~al.}(2016){Jin}, {Li}, {Isella}, {Li}, \& {Ji}}]{jin2016}
{Jin}, S., {Li}, S., {Isella}, A., {Li}, H., \& {Ji}, J. 2016, \apj, 818, 76

\bibitem[{{Kenyon} \& {Hartmann}(1987)}]{kenyon1987}
{Kenyon}, S.~J. \& {Hartmann}, L. 1987, \apj, 323, 714

\bibitem[{{Kirchschlager} {et~al.}(2016){Kirchschlager}, {Wolf}, \&
  {Madlener}}]{Kirchschlager2016}
{Kirchschlager}, F., {Wolf}, S., \& {Madlener}, D. 2016, \mnras, 462, 858

\bibitem[{{Kirkpatrick} {et~al.}(1983){Kirkpatrick}, {Gelatt}, \&
  {Vecchi}}]{kirkpatrick1983}
{Kirkpatrick}, S., {Gelatt}, C.~D., \& {Vecchi}, M.~P. 1983, Science, 220, 671

\bibitem[{{Knigge}(1999)}]{knigge1999}
{Knigge}, C. 1999, \mnras, 309, 409

\bibitem[{{Kurucz}(1994)}]{Kurucz1994}
{Kurucz}, R. 1994, Solar abundance model atmospheres for 0,1,2,4,8 km/s.~Kurucz
  CD-ROM No.~19.~ Cambridge, Mass.: Smithsonian Astrophysical Observatory,
  1994., 19

\bibitem[{{Kwon} {et~al.}(2011){Kwon}, {Looney}, \& {Mundy}}]{kwon2011}
{Kwon}, W., {Looney}, L.~W., \& {Mundy}, L.~G. 2011, \apj, 741, 3

\bibitem[{{Lada}(1987)}]{lada1987}
{Lada}, C.~J. 1987, in IAU Symposium, Vol. 115, Star Forming Regions, ed.
  M.~{Peimbert} \& J.~{Jugaku}, 1--17

\bibitem[{{Lissauer} \& {Stevenson}(2007)}]{lissauer2007}
{Lissauer}, J.~J. \& {Stevenson}, D.~J. 2007, Protostars and Planets V, 591

\bibitem[{{Liu} {et~al.}(2012){Liu}, {Madlener}, {Wolf}, {Wang}, \&
  {Ruge}}]{lium2012}
{Liu}, Y., {Madlener}, D., {Wolf}, S., {Wang}, H., \& {Ruge}, J.~P. 2012, \aap,
  546, A7

\bibitem[{{Madlener} {et~al.}(2012){Madlener}, {Wolf}, {Dutrey}, \&
  {Guilloteau}}]{madlener2012}
{Madlener}, D., {Wolf}, S., {Dutrey}, A., \& {Guilloteau}, S. 2012, \aap, 543,
  A81

\bibitem[{{Men'shchikov} {et~al.}(1999){Men'shchikov}, {Henning}, \&
  {Fischer}}]{menshchikov1999}
{Men'shchikov}, A.~B., {Henning}, T., \& {Fischer}, O. 1999, \apj, 519, 257

\bibitem[{{Mundt} {et~al.}(1990){Mundt}, {Buehrke}, {Solf}, {Ray}, \&
  {Raga}}]{mundt1990}
{Mundt}, R., {Buehrke}, T., {Solf}, J., {Ray}, T.~P., \& {Raga}, A.~C. 1990,
  \aap, 232, 37

\bibitem[{{Natta} {et~al.}(2007){Natta}, {Testi}, {Calvet}, {Henning},
  {Waters}, \& {Wilner}}]{natta2007}
{Natta}, A., {Testi}, L., {Calvet}, N., {et~al.} 2007, Protostars and Planets
  V, 767

\bibitem[{{Okuzumi} {et~al.}(2016){Okuzumi}, {Momose}, {Sirono}, {Kobayashi},
  \& {Tanaka}}]{okuzumi2016}
{Okuzumi}, S., {Momose}, M., {Sirono}, S.-i., {Kobayashi}, H., \& {Tanaka}, H.
  2016, \apj, 821, 82

\bibitem[{{Papaloizou} \& {Terquem}(2006)}]{papaloizou2006}
{Papaloizou}, J.~C.~B. \& {Terquem}, C. 2006, Reports on Progress in Physics,
  69, 119

\bibitem[{{P{\'e}rez} {et~al.}(2012){P{\'e}rez}, {Carpenter}, {Chandler},
  {Isella}, {Andrews}, {Ricci}, {Calvet}, {Corder}, {Deller}, {Dullemond},
  {Greaves}, {Harris}, {Henning}, {Kwon}, {Lazio}, {Linz}, {Mundy}, {Sargent},
  {Storm}, {Testi}, \& {Wilner}}]{perez2012}
{P{\'e}rez}, L.~M., {Carpenter}, J.~M., {Chandler}, C.~J., {et~al.} 2012,
  \apjl, 760, L17

\bibitem[{{P{\'e}rez} {et~al.}(2015){P{\'e}rez}, {Chandler}, {Isella},
  {Carpenter}, {Andrews}, {Calvet}, {Corder}, {Deller}, {Dullemond}, {Greaves},
  {Harris}, {Henning}, {Kwon}, {Lazio}, {Linz}, {Mundy}, {Ricci}, {Sargent},
  {Storm}, {Tazzari}, {Testi}, \& {Wilner}}]{perez2015}
{P{\'e}rez}, L.~M., {Chandler}, C.~J., {Isella}, A., {et~al.} 2015, \apj, 813,
  41

\bibitem[{{Picogna} \& {Kley}(2015)}]{picogna2015}
{Picogna}, G. \& {Kley}, W. 2015, \aap, 584, A110

\bibitem[{{Pinte} {et~al.}(2016){Pinte}, {Dent}, {M{\'e}nard}, {Hales}, {Hill},
  {Cortes}, \& {de Gregorio-Monsalvo}}]{pinte2016}
{Pinte}, C., {Dent}, W.~R.~F., {M{\'e}nard}, F., {et~al.} 2016, \apj, 816, 25

\bibitem[{{Pinte} {et~al.}(2007){Pinte}, {Fouchet}, {M{\'e}nard}, {Gonzalez},
  \& {Duch{\^e}ne}}]{pinte2007}
{Pinte}, C., {Fouchet}, L., {M{\'e}nard}, F., {Gonzalez}, J.-F., \&
  {Duch{\^e}ne}, G. 2007, \aap, 469, 963

\bibitem[{{Pinte} {et~al.}(2008){Pinte}, {Padgett}, {M{\'e}nard},
  {Stapelfeldt}, {Schneider}, {Olofsson}, {Pani{\'c}}, {Augereau},
  {Duch{\^e}ne}, {Krist}, {Pontoppidan}, {Perrin}, {Grady}, {Kessler-Silacci},
  {van Dishoeck}, {Lommen}, {Silverstone}, {Hines}, {Wolf}, {Blake}, {Henning},
  \& {Stecklum}}]{pinte2008}
{Pinte}, C., {Padgett}, D.~L., {M{\'e}nard}, F., {et~al.} 2008, \aap, 489, 633

\bibitem[{{Podolak} \& {Zucker}(2004)}]{podolak2004}
{Podolak}, M. \& {Zucker}, S. 2004, Meteoritics and Planetary Science, 39, 1859

\bibitem[{{Pohl} {et~al.}(2015){Pohl}, {Pinilla}, {Benisty}, {Ataiee},
  {Juh{\'a}sz}, {Dullemond}, {Van Boekel}, \& {Henning}}]{pohl2015}
{Pohl}, A., {Pinilla}, P., {Benisty}, M., {et~al.} 2015, \mnras, 453, 1768

\bibitem[{{Pontoppidan} {et~al.}(2014){Pontoppidan}, {Salyk}, {Bergin},
  {Brittain}, {Marty}, {Mousis}, \& {{\"O}berg}}]{pontoppidan2014}
{Pontoppidan}, K.~M., {Salyk}, C., {Bergin}, E.~A., {et~al.} 2014, Protostars
  and Planets VI, 363

\bibitem[{{Rebull} {et~al.}(2004){Rebull}, {Wolff}, \& {Strom}}]{rebull2004}
{Rebull}, L.~M., {Wolff}, S.~C., \& {Strom}, S.~E. 2004, \aj, 127, 1029

\bibitem[{{Ricci} {et~al.}(2010{\natexlab{a}}){Ricci}, {Testi}, {Natta}, \&
  {Brooks}}]{ricci2010b}
{Ricci}, L., {Testi}, L., {Natta}, A., \& {Brooks}, K.~J. 2010{\natexlab{a}},
  \aap, 521, A66

\bibitem[{{Ricci} {et~al.}(2010{\natexlab{b}}){Ricci}, {Testi}, {Natta},
  {Neri}, {Cabrit}, \& {Herczeg}}]{ricci2010a}
{Ricci}, L., {Testi}, L., {Natta}, A., {et~al.} 2010{\natexlab{b}}, \aap, 512,
  A15

\bibitem[{{Robitaille} {et~al.}(2007){Robitaille}, {Whitney}, {Indebetouw}, \&
  {Wood}}]{robitaille2007}
{Robitaille}, T.~P., {Whitney}, B.~A., {Indebetouw}, R., \& {Wood}, K. 2007,
  \apjs, 169, 328

\bibitem[{{Ruge} {et~al.}(2016){Ruge}, {Flock}, {Wolf}, {Dzyurkevich},
  {Fromang}, {Henning}, {Klahr}, \& {Meheut}}]{ruge2016}
{Ruge}, J.~P., {Flock}, M., {Wolf}, S., {et~al.} 2016, \aap, 590, A17

\bibitem[{{Sauter} {et~al.}(2009){Sauter}, {Wolf}, {Launhardt}, {Padgett},
  {Stapelfeldt}, {Pinte}, {Duch{\^e}ne}, {M{\'e}nard}, {McCabe}, {Pontoppidan},
  {Dunham}, {Bourke}, \& {Chen}}]{sauter2009}
{Sauter}, J., {Wolf}, S., {Launhardt}, R., {et~al.} 2009, \aap, 505, 1167

\bibitem[{{Takahashi} \& {Inutsuka}(2014)}]{takahashi2014}
{Takahashi}, S.~Z. \& {Inutsuka}, S.-i. 2014, \apj, 794, 55

\bibitem[{{Takahashi} \& {Inutsuka}(2016)}]{takahashi2016}
{Takahashi}, S.~Z. \& {Inutsuka}, S.-i. 2016, ArXiv e-prints

\bibitem[{{Tazzari} {et~al.}(2016){Tazzari}, {Testi}, {Ercolano}, {Natta},
  {Isella}, {Chandler}, {P{\'e}rez}, {Andrews}, {Wilner}, {Ricci}, {Henning},
  {Linz}, {Kwon}, {Corder}, {Dullemond}, {Carpenter}, {Sargent}, {Mundy},
  {Storm}, {Calvet}, {Greaves}, {Lazio}, \& {Deller}}]{tazzari2016}
{Tazzari}, M., {Testi}, L., {Ercolano}, B., {et~al.} 2016, \aap, 588, A53

\bibitem[{{Testi} {et~al.}(2014){Testi}, {Birnstiel}, {Ricci}, {Andrews},
  {Blum}, {Carpenter}, {Dominik}, {Isella}, {Natta}, {Williams}, \&
  {Wilner}}]{testi2014}
{Testi}, L., {Birnstiel}, T., {Ricci}, L., {et~al.} 2014, Protostars and
  Planets VI, 339

\bibitem[{{Testi} {et~al.}(2015){Testi}, {Skemer}, {Henning}, {Bailey},
  {Defr{\`e}re}, {Hinz}, {Leisenring}, {Vaz}, {Esposito}, {Fontana}, {Marconi},
  {Skrutskie}, \& {Veillet}}]{testi2015}
{Testi}, L., {Skemer}, A., {Henning}, T., {et~al.} 2015, \apjl, 812, L38

\bibitem[{{Trotta} {et~al.}(2013){Trotta}, {Testi}, {Natta}, {Isella}, \&
  {Ricci}}]{trotta2013}
{Trotta}, F., {Testi}, L., {Natta}, A., {Isella}, A., \& {Ricci}, L. 2013,
  \aap, 558, A64

\bibitem[{{Ward}(2000)}]{ward2000}
{Ward}, W.~R. 2000, {On Planetesimal Formation: The Role of Collective Particle
  Behavior}, ed. R.~M. {Canup}, K.~{Righter}, \& {et al.}, 75--84

\bibitem[{{White} \& {Hillenbrand}(2004)}]{white2004}
{White}, R.~J. \& {Hillenbrand}, L.~A. 2004, \apj, 616, 998

\bibitem[{{Wolf} {et~al.}(2003){Wolf}, {Padgett}, \& {Stapelfeldt}}]{wolfp2003}
{Wolf}, S., {Padgett}, D.~L., \& {Stapelfeldt}, K.~R. 2003, \apj, 588, 373

\bibitem[{{Yen} {et~al.}(2016){Yen}, {Liu}, {Gu}, {Hirano}, {Lee},
  {Puspitaningrum}, \& {Takakuwa}}]{yen2016}
{Yen}, H.-W., {Liu}, H.~B., {Gu}, P.-G., {et~al.} 2016, \apjl, 820, L25

\bibitem[{{Youdin}(2011)}]{youdin2011}
{Youdin}, A.~N. 2011, \apj, 731, 99

\bibitem[{{Zhang} {et~al.}(2015){Zhang}, {Blake}, \& {Bergin}}]{zhang2015}
{Zhang}, K., {Blake}, G.~A., \& {Bergin}, E.~A. 2015, \apjl, 806, L7

\end{thebibliography}

\end{document}